\documentclass[twocolumn,showpacs,preprintnumbers,prl]{revtex4}

\usepackage{graphicx}
\usepackage{dcolumn}
\usepackage{bm}
\usepackage{epsfig}
\usepackage[usenames]{color}
\begin{document}


\title{Single photo-electron trapping, storage, and detection in a one-electron quantum dot}

\author{Deepak Sethu Rao}
\author{Thomas Szkopek}
\author{Hans Daniel Robinson}
\author{Eli Yablonovitch}

\affiliation{Electrical Engineering Department, University of
California Los Angeles, Los Angeles, CA 90095}

\author{Hong-Wen Jiang}
 \affiliation {Department of Physics and Astronomy, University of California Los
 Angeles, Los Angeles, CA 90095}

\date{\today}

\begin{abstract}
There has been considerable progress in electro-statically
emptying, and re-filling, quantum dots with individual electrons.
Typically the quantum dot is defined by electrostatic gates on a
GaAs/Al$_\mathrm{y}$Ga$_\mathrm{{1-y}}$As modulation doped
heterostructure. We report the filling of such a quantum dot by a
single photo-electron, originating from an individual photon. The
electrostatic dot can be emptied and reset in a controlled fashion
before the arrival of each photon. The trapped photo-electron is
detected by a point contact transistor integrated adjacent to the
electrostatic potential trap. Each stored photo-electron causes a
persistent negative step in the transistor channel current.  Such
a controllable, benign, single photo-electron detector could allow
for information transfer between flying photon qubits and stored
electron qubits.
\end{abstract}


\maketitle

The detection of a single photo-electron generally requires some
type of gain mechanism.  A new mechanism has emerged recently,
photoconductive gain \cite{Rose}, for providing the sensitivity
needed for single charge detection
\cite{Shields2000,Kosaka1,Kosaka2}. Indeed, the detection of a
photo-hole is easier and more common than the detection of a
photo-electron. The positive charge of a trapped photo-hole
attracts electrons and leads to conventional positive
photo-conductivity. Recently, single photon detection has been
demonstrated by photo-hole trapping in defects \cite{Kosaka1} and
self-assembled quantum dots \cite{Shields2000} within
semiconductors. The trapping of a photo-electron on the otherhand
repels current, and leads to the more exotic \cite{Rose}
``negative photoconductivity''. Photo-electron trapping has thus
far been demonstrated in the microwave regime by photon-assisted
tunnelling between Landau levels \cite{Komiyama1} and in an
electrostatic quantum dot \cite{Kosaka2} with limited or no
control over systematic emptying and injecting a single
photo-electron. In this paper we report the trapping and detection
of a single, inter-band photo-electron in a controllable
electrostatic quantum dot.

The benefit of safely and gently trapping a photo-electron is that
its spin information may be preserved.  Favorable selection rules
for information transfer between quantum states of photons and
spin states of electrons in semiconductors have been identified
\cite{Rutger1}.  It may become possible to transfer quantum
information over long distances by exchanging information between
flying qubits and stationary qubits \cite{Bennett-93}.

\begin{figure}
\centering
 \epsfig{file=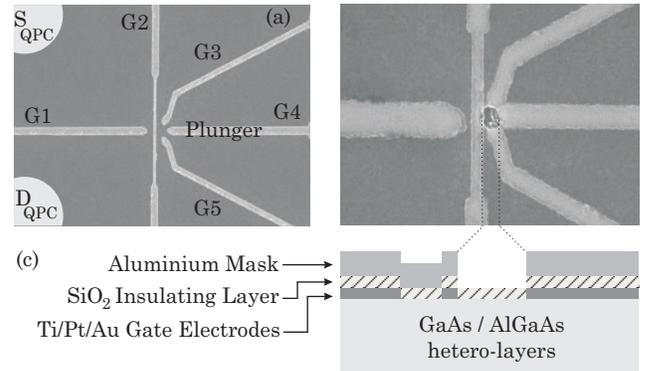,width=8.4cm,height=5.0cm}
 \caption{ (a) Scanning Electron Micrograph of the surface metallic gates
 defining a Quantum Point Contact between the Source and Drain ohmic contacts
 (S$_{\mathrm{QPC}}$ \& D$_{\mathrm{QPC}}$) and a lateral electrostatic Quantum Dot.  (b) SEM
 of pin-hole aperture etched in an opaque Al layer, 150nm thick,
acting as a shadow mask to illuminate only the Quantum Dot region.
Gates are buried under Al/SiO$_2$ layers (c) Cross-section view of
the device. The GaAs/AlGaAs hetero-layers consist of a 5nm
Si-doped (1x10$^{18}$/cm$^3$) GaAs cap layer, a 60nm Si-doped
(1x10$^{18}$/cm$^3$) n-Al$_{0.3}$Ga$_{0.7}$As layer, a 30nm
i-Al$_{0.3}$Ga$_{0.7}$As spacer layer, on an undoped GaAs buffer.}
 \label{fig:SEM}
\end{figure}

It is essential that any new opto-spintronic device designed to
achieve the above objectives accomplishes the following tasks: (i)
trap a photo-excited electron in an artificially engineered trap;
(ii) detect the stored electron by means of a benign gain
mechanism; and most importantly (iii) ensure that the trap holds
none but the single photo-excited electron.  We experimentally
demonstrate the injection and detection of a single, inter-band
photo-excited electron, into an empty quantum dot defined
electrostatically by metallic gates on a GaAs/AlGaAs
heterostructure, with an integrated charge read-out transistor.

The signature of photo-electron trapping is negative
photo-conductivity - a drop in the current through the detection
circuit upon illumination \cite{Kosaka2}.  Negative photo
conductivity is commonly not observed in GaAs/AlGaAs
hetrostructures, though a persistent photo-induced increase in
conductivity has been well known for some time now \cite{Nelson}.
Positive photo-conductivity is a result of the trapping of
photo-excited holes and a subsequent increase in the 2D electrons
gas (2DEG) density.  We have earlier reported the detection of
such individual photo-hole trapping events with a simple
split-gate geometry \cite{Kosaka1}. Photo-holes are trapped
predominantly by negatively charged defects at low temperatures
known as DX centers.  Persistent negative photo-conductivity at
low temperatures has been reported only after the saturation of
hole trapping centers, most likely ionized donors, and only at
short wavelengths causing photo-excitation in the doped AlGaAs
barrier layer \cite{Klitzing-89,Wilson}.

On the other hand, photo-excitation in GaAs has always shown a
positive increase in conductivity.  Now, by creating an artificial
electron trap defined by electrostatic metal gate electrodes, we
have been able to detect the addition of a single photo-excited
electron into the electron trap.  We suppressed the usually
dominant positive photo-conductivity by a shadow mask, that
permitted the light to fall only in the immediate vicinity of the
electrostatic quantum dot.  A point contact field-effect
transistor integrated adjacent to the dot \cite{Field} serves to
detect the injected photo-electron in a non-intrusive way. We
believe that sensing a current directly through the quantum dot
would be too invasive.

Our device is fabricated on a modulation doped
GaAs/Al$_{0.3}$Ga$_{0.7}$As heterostructure grown by molecular
beam epitaxy on a semi-insulating GaAs substrate. A scanning
electron micrograph of the gate geometry of the device used in our
measurements is shown in Fig.\protect\ref{fig:SEM}. The gates are
fabricated by electron beam lithography and electron-gun
evaporation of Ti/Pt/Au. G1 and G2 define a quantum point contact
(QPC) between the left source and drain Ohmic contacts,
S$_\mathrm{{QPC}}$ and D$_\mathrm{{QPC}}$ respectively, shown in
Fig.\protect\ref{fig:SEM}(a). Adjacent to this point contact, an
electrostatic circular quantum dot with a lithographic radius of
200nm is defined by gates G3, G4 and G5. The electrostatic dot is
defined by squeezing the 2DEG by the surface metallic gates.  A
variety of experiments have studied the properties of such
GaAs/AlGaAs quantum dots in great detail
\cite{Field,Ciorga,Shtrikman1,Kouwen2003-1}, and a vast knowledge
base has been developed.

\begin{figure}
\centering
 \epsfig{file=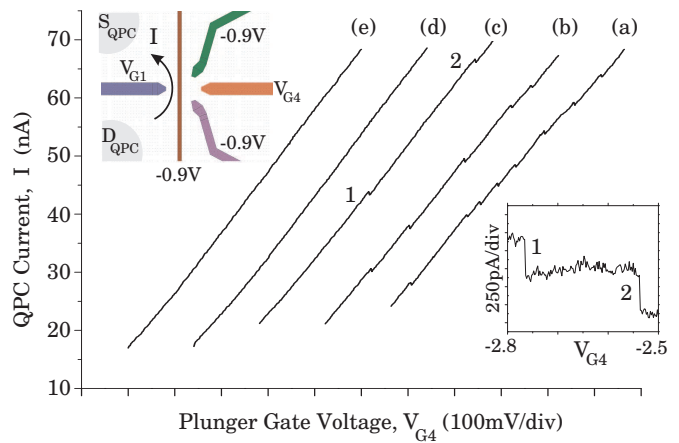,width=8.75cm,height=5.75cm}
 \caption{Single electron escape from the dot detected by the QPC
transistor.  The plunger is swept from -1.5V to -4V with a scan
rate of 4mV/sec starting at curve marked (a) and ending at (e)
with each curve spanning 500mV.
 In-between each curve, V$_{\mathrm{G1}}$ is changed to reset the QPC current.
 The curves have been offset along the voltage axis for clarity. The top inset shows
 the operating voltages on all the gates. The bottom inset shows the step sizes of
 the last two electrons in the dot seen in curve (c) after subtracting out the background slope
 (V$_{\mathrm{SD}}$(QPC)=3.25mV).}
 \label{fig:IVSlow}
\end{figure}

Negative voltages on the five surface gates isolate a puddle of
electrons in the 2DEG adjacent to the point contact transistor.
Gates G3 and G5 together with G2 control the tunnel coupling of
the electrons in the dot to the external 2DEG reservoirs, while
gate G4 is used as a plunger to push electrons out of the dot one
at a time down to the last electron.  This creates an empty dot
just before exposure to light.  Photo-events over the bulk of the
device are suppressed by a 150nm thick Aluminium layer deposited
as a mask over the entire area of the device, except for a
pin-hole aperture directly above the quantum dot as shown in the
SEM of Fig.\protect\ref{fig:SEM}(b).  An insulating SiO$_2$ layer
and a thin adhesion layer of Titanium separate the metal gate
electrodes from the Aluminium mask layer.
Fig.\protect\ref{fig:SEM}(c) shows the cross section view of the
device layers.

\begin{figure}
\centering
 \epsfig{file=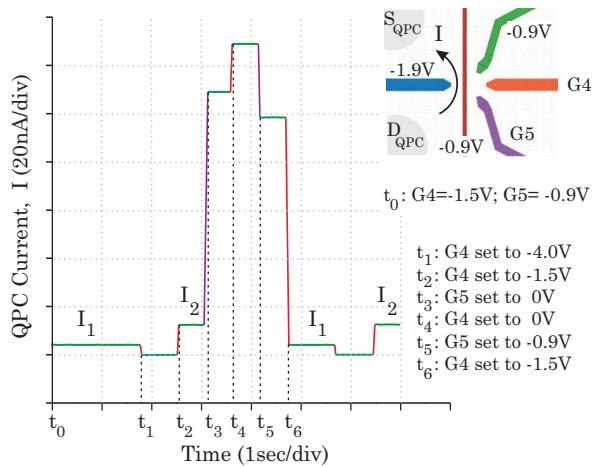,width=7.75cm,height=6.15cm}
 \caption{Hysteresis measured in the current through the QPC transistor,
 associated with the transition of the dot from the meta-stable filled state to the equilibrium empty state.
 The current switches from I$_1$ to I$_2$ as the G4 plunger gate ejects stored electrons
 in the cycle from t$_0$ to t$_2$.  When the barriers are re-opened and closed in the cycle from
 t$_3$ to t$_6$, electrons remain trapped in the dot restoring the current to I$_1$.
 The color of the vertical transitions is coded to the color of the
 corresponding gate switch for that transition.  Level I$_2$ represents the
 desired empty state of the dot, at which it is ready to accept and trap photo-injected electrons.}
 \label{fig:BiStable}
\end{figure}

We first present the electrical characterization of the
electrostatic dot in Fig.\protect\ref{fig:IVSlow} and
Fig.\protect\ref{fig:BiStable}. The device is cooled gradually to
0.43K in a $^3$He cryostat and negative voltages are applied to
the five metallic surface gates defining the dot and the QPC.
Fig.\protect\ref{fig:IVSlow} plots the current through the point
contact transistor versus the plunger gate voltage,
V$_\mathrm{{G4}}$. The plunger is swept at a rate of 4mV/sec to
repel electrons one at a time, into the surrounding 2DEG. It is
important to detect the single electron tunnelling events and the
trapped electric charge by means of an adjacent transistor, rather
than by invasively passing current through the dot storing the
electron. As an electron escapes the electrostatic quantum dot,
the diminished electrostatic repulsion causes a jump in the QPC
transistor current.

The quantum dot state at the start of the scan in
Fig.\protect\ref{fig:IVSlow} is the same as that at time t$_0$,
(or equivalently t$_6$) in Fig.\protect\ref{fig:BiStable}. Upon
formation, a few excess electrons remain trapped in the dot in a
long lived meta-stable state, prior to being forced out by the
plunger gate. The point contact current varies in a saw-tooth
fashion with a small discrete positive step for each electron
ejected as seen in Fig.\protect\ref{fig:IVSlow}. The last electron
emission event occurs on curve (c) at a voltage of about G4=-2.75V
on the plunger gate. In order to ensure that the absence of
further steps is not due to very slow tunnelling times, the
barrier gate voltage G3 was raised just after the last detected
step to allow any remaining electrons to escape. Only a smooth
increase in the QPC current could be observed due to the
capacitive coupling between the point contact and the tunnel
barrier gate with no evidence for any remaining electrons.
Electron tunnelling from the dot in this regime is essentially a
statistical process, but it is sped up according the G3, G4, G5
gate voltage settings. The lower inset to
Fig.\protect\ref{fig:IVSlow} shows the steps corresponding to the
last two electrons after subtracting out the background slope.
Close to the optimum sensitivity point of source/drain
conductance, {1/R}$_{\mathrm{QPC}}\sim$0.5(2e$^2$/h), the observed
single electron step is 0.5nA providing an excellent signal to
noise ratio.

Upon sweeping the plunger gate G4 back up to -1.5V from -4V at the
same scan rate as in the forward direction, no electrons were
observed to re-enter the dot. The equilibrium state of the dot at
V$_\mathrm{{G4}}$ = -1.5V is the ``empty dot" state, since by that
point, the dot energy levels have been raised well above the
external Fermi level. Fig.\protect\ref{fig:BiStable} illustrates
the hysteretic behavior of the QPC transistor source/drain current
associated with the emission of electrons from the dot.
Immediately following time t$_0$, and equivalently time t$_6$, the
dot exists in the meta-stable state with excess trapped electrons.
The thick tunnel barriers formed in our geometry when G3 and G5
are at -0.9V prevent fast tunnelling. No electrons were observed
to escape in the interval between time t$_0$ and t$_1$, while at
t$_1$ they are forced out. The electron emission associated with
the slow 5 minutes ramp in Fig.\protect\ref{fig:IVSlow} occurs
within the plunger rise-time at t$_1$ in
Fig.\protect\ref{fig:BiStable}. The gate voltage changes at
t$_1$,t$_2$,t$_3$,t$_4$,t$_5$ all lead to an equilibrium electron
density. The QPC current level I$_1$ represents the filled
meta-stable dot state and level I$_2$ the empty state, at which
the dot is ready to accept and hold only the photo-injected
electron with a storage time longer than 5 minutes.

Highly attenuated light pulses at a vacuum wavelength
$\lambda$=760nm, which photo-excite inter-band electrons in the
GaAs layer, were created by a Pockels cell modulator at the output
of a cw laser. The pulses were focused onto a spot size of about
100$\mathrm{{\mu}}$m diameter on the sample. The opaque Aluminium
mask blocks almost all of the incident photon flux except directly
above the electrostatic dot where there was a 200nm radius
pin-hole aperture. Assuming a Gaussian profile for the incident
spot over the illumination area of radius 50$\mathrm{\mu}$m, and
given the 200nm radius of the electrostatic dot, the photon flux
into the dot is reduced by a factor of 10$^{-5}$ compared to the
total incident flux.

\begin{figure}
\centering
 \epsfig{file=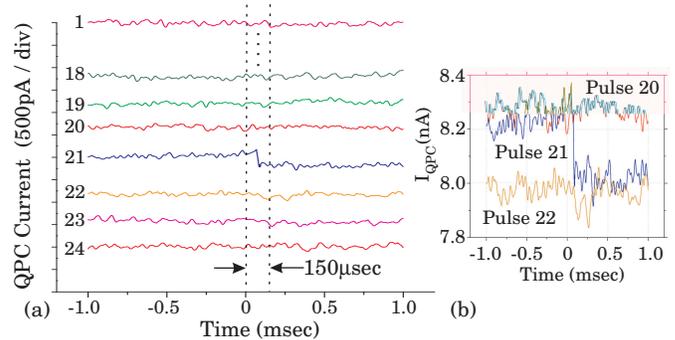,width=8.75cm,height=4.5cm}
 \caption{ (a) Photo-electron trapping in the quantum dot detected by adjacent
 point contact transistor. The dot is fully emptied before exposure to
 $\lambda$=760nm pulses, at a flux of 0.1photons/pulse into the dot, within a 150$\mathrm{\mu}$sec time window.
 The time traces depict the transistor current, centered on the
pulse time window. The traces have been offset for clarity. (b) An
expanded view of transistor current for pulses 20,21 and 22
without any offset.  The charge sensitivity per photo-electron is
10$^{-3}$e/$\sqrt\mathrm{{Hz}}$.}
 \label{fig:NStep}
\end{figure}

Fig.\protect\ref{fig:NStep}, which plots the QPC transistor
current versus time, presents a typical experimental result of
exposure to a series of consecutive pulses after emptying the dot,
prior to the first pulse. In this figure, the incident photon flux
was maintained at 0.1 photons/pulse within the dot area. Time t=0
marks the time at which the Pockels cell was opened, for a pulse
duration of 150$\mathrm{\mu}$sec. When a photon is absorbed within
the active area, and the photo electron gets trapped in the dot, a
sharp drop in transistor current is seen for pulse 21 in the
series. The current step size is consistent with the expected
single electron steps determined from the electrical
characterization in Fig.\protect\ref{fig:IVSlow}. After emptying
the dot by the plunger gate G4, if even any one of the gates G3,G4
or G5 is grounded, the quantum dot is open and negative
photo-conductivity steps were not observed. We thus rule out the
possibility of negative photo-conductucivity steps due
photo-electron trapping in donors, DX-centers and traps in the
SiO$_2$ layer. The fall time associated with the single electron
signal is 20$\mathrm{\mu}$sec, from Fig.
\protect\ref{fig:NStep}(b), consistent with the speed of the
pre-amp that was used. Given the signal-to-noise ratio in Fig.
\protect\ref{fig:NStep}(b), this leads to a single photo-electron
signal to noise ratio of about 10$^{-3}$
electrons/$\sqrt\mathrm{{Hz}}$.

\begin{figure}
\centering
 \epsfig{file=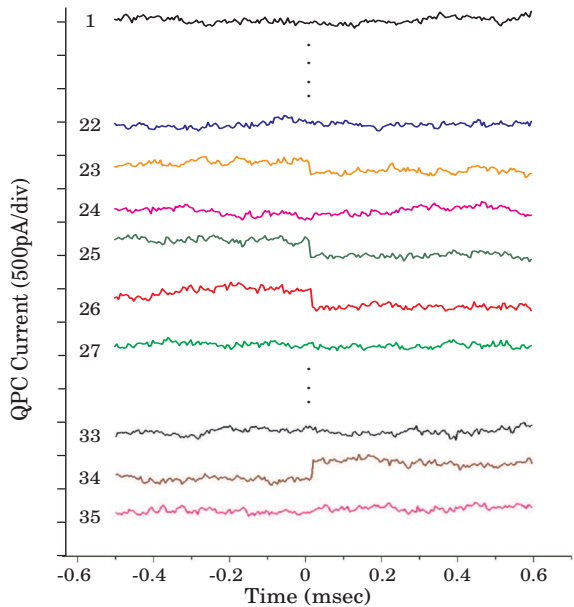,width=7.5cm,height=8cm}
 \caption{An optical pulse series with an average flux of
 1.2photons/pulse within the dot area.  Occasional positive steps can be
 attributed to the photo-ionization of a residual neutral donor, or the annihilation of a photo-hole within the electrostatic dot.}
 \label{fig:NPStep}
\end{figure}

Increasing the photon flux over the dot increases the frequency of
occurrence of negative photo-conductivity steps.
Fig.\protect\ref{fig:NPStep} shows a series of traces for a photon
flux of 1.2 photons/pulse into the dot with no reset to empty the
dot between pulses. Based on the frequency of occurrence of
photo-detection events, we estimate the photo-electron trapping
quantum efficiency to be about 10\%. This is consistent with the
penetration depth of $\lambda$=760nm light, and the size of the
electrostatic potential dot. Interspersed among the negative
steps, some positive steps were occasionally seen, as in the
34$^\mathrm{th}$ pulse in Fig.\protect\ref{fig:NPStep}. Such
positive signals were seen with a 1\% occurrence rate and can be
attributed to the photo-ionization of residual neutral donors
close to the quantum dot. The occasional positive steps were more
noticeable when the dot held several photo-electrons, possibly due
to the additional mechanism of photo-electron ionization or
photo-hole annihilation within the dot. The positive steps are
rare since almost all the photo-holes are swept away by the
surrounding negatively biased gate electrodes.

In conclusion, we have demonstrated single photo-electron trapping
and storage in an empty electrostatic quantum dot that can be
controllably created prior to photo-excitation of inter-band
electrons. Recently, experiments demonstrating the electrical
measurement of a single electron spin inserted in a similar
electrostatic dot \cite{Kouwen2004-1} or in a commercial Si
field-effect transistor \cite{MingX} have been reported. The
successful trapping and detection of photo-electrons reported
here, in spite of the usually dominant positive
photo-conductivity, would enable the implementation of a detector
for an optically injected spin. By combining the single
photo-electron trapping result reported in this paper, with the
single spin measurement reported in \cite{Kouwen2004-1}, it would
be possible to convert a flying qubit (photon) into a stationary
qubit (trapped electron) and to measure the spin state.

The work is supported by the Defense Advanced Research Projects
Agency (MDA972-99-1-0017), Army Research Office (DAAD19-00-1-0172)
\& the Defense MicroElectronics Activity.


\begin{thebibliography}{16}
\expandafter\ifx\csname
natexlab\endcsname\relax\def\natexlab#1{#1}\fi
\expandafter\ifx\csname bibnamefont\endcsname\relax
  \def\bibnamefont#1{#1}\fi
\expandafter\ifx\csname bibfnamefont\endcsname\relax
  \def\bibfnamefont#1{#1}\fi
\expandafter\ifx\csname citenamefont\endcsname\relax
  \def\citenamefont#1{#1}\fi
\expandafter\ifx\csname url\endcsname\relax
  \def\url#1{\texttt{#1}}\fi
\expandafter\ifx\csname
urlprefix\endcsname\relax\def\urlprefix{URL }\fi
\providecommand{\bibinfo}[2]{#2}
\providecommand{\eprint}[2][]{\url{#2}}

\bibitem[{\citenamefont{Rose}(1978)}]{Rose}
\bibinfo{author}{\bibfnamefont{A.}~\bibnamefont{Rose}},
  \emph{\bibinfo{title}{Concepts in Photoconductivity and Allied Problems}}
  (\bibinfo{publisher}{Krieger Publishing Co.}, \bibinfo{address}{Huntington,
  New York}, \bibinfo{year}{1978}).

\bibitem[{\citenamefont{Shields et~al.}(2000)\citenamefont{Shields, O'Sullivan,
  Farrer, Ritchie, Hogg, Leadbeater, Norman, and Pepper}}]{Shields2000}
\bibinfo{author}{\bibfnamefont{A.}~\bibnamefont{Shields}},
  \bibinfo{author}{\bibfnamefont{M.}~\bibnamefont{O'Sullivan}},
  \bibinfo{author}{\bibfnamefont{I.}~\bibnamefont{Farrer}},
  \bibinfo{author}{\bibfnamefont{D.}~\bibnamefont{Ritchie}},
  \bibinfo{author}{\bibfnamefont{R.}~\bibnamefont{Hogg}},
  \bibinfo{author}{\bibfnamefont{M.}~\bibnamefont{Leadbeater}},
  \bibinfo{author}{\bibfnamefont{C.}~\bibnamefont{Norman}}, \bibnamefont{and}
  \bibinfo{author}{\bibfnamefont{M.}~\bibnamefont{Pepper}},
  \bibinfo{journal}{Appl.\ Phys.\ Lett.} \textbf{\bibinfo{volume}{76}},
  \bibinfo{pages}{3673} (\bibinfo{year}{2000}).

\bibitem[{\citenamefont{Kosaka et~al.}(2002)\citenamefont{Kosaka, Rao,
  Robinson, Bandaru, Sakamoto, and Yablonovitch}}]{Kosaka1}
\bibinfo{author}{\bibfnamefont{H.}~\bibnamefont{Kosaka}},
  \bibinfo{author}{\bibfnamefont{D.~S.} \bibnamefont{Rao}},
  \bibinfo{author}{\bibfnamefont{H.~D.} \bibnamefont{Robinson}},
  \bibinfo{author}{\bibfnamefont{P.}~\bibnamefont{Bandaru}},
  \bibinfo{author}{\bibfnamefont{T.}~\bibnamefont{Sakamoto}}, \bibnamefont{and}
  \bibinfo{author}{\bibfnamefont{E.}~\bibnamefont{Yablonovitch}},
  \bibinfo{journal}{Phys.\ Rev.\ B} \textbf{\bibinfo{volume}{65}},
  \bibinfo{pages}{R2013071} (\bibinfo{year}{2002}).

\bibitem[{\citenamefont{Kosaka et~al.}(2003)\citenamefont{Kosaka, Rao,
  Robinson, Bandaru, Makita, and Yablonovitch}}]{Kosaka2}
\bibinfo{author}{\bibfnamefont{H.}~\bibnamefont{Kosaka}},
  \bibinfo{author}{\bibfnamefont{D.~S.} \bibnamefont{Rao}},
  \bibinfo{author}{\bibfnamefont{H.~D.} \bibnamefont{Robinson}},
  \bibinfo{author}{\bibfnamefont{P.}~\bibnamefont{Bandaru}},
  \bibinfo{author}{\bibfnamefont{K.}~\bibnamefont{Makita}}, \bibnamefont{and}
  \bibinfo{author}{\bibfnamefont{E.}~\bibnamefont{Yablonovitch}},
  \bibinfo{journal}{Phys.\ Rev.\ B} \textbf{\bibinfo{volume}{67}},
  \bibinfo{pages}{045104} (\bibinfo{year}{2003}).

\bibitem[{\citenamefont{Komiyama et~al.}(2000)\citenamefont{Komiyama, Astafiev,
  Antonov, T.Kutsuwa, and H.Hirai}}]{Komiyama1}
\bibinfo{author}{\bibfnamefont{S.}~\bibnamefont{Komiyama}},
  \bibinfo{author}{\bibfnamefont{O.}~\bibnamefont{Astafiev}},
  \bibinfo{author}{\bibfnamefont{V.}~\bibnamefont{Antonov}},
  \bibinfo{author}{\bibnamefont{T.Kutsuwa}}, \bibnamefont{and}
  \bibinfo{author}{\bibnamefont{H.Hirai}}, \bibinfo{journal}{Nature}
  \textbf{\bibinfo{volume}{403}}, \bibinfo{pages}{405} (\bibinfo{year}{2000}).

\bibitem[{\citenamefont{Vrijen and Yablonovitch}(2001)}]{Rutger1}
\bibinfo{author}{\bibfnamefont{R.}~\bibnamefont{Vrijen}} \bibnamefont{and}
  \bibinfo{author}{\bibfnamefont{E.}~\bibnamefont{Yablonovitch}},
  \bibinfo{journal}{Physica \ E \ (Amsterdam)} \textbf{\bibinfo{volume}{10}},
  \bibinfo{pages}{569} (\bibinfo{year}{2001}).

\bibitem[{\citenamefont{Bennett et~al.}(1993)\citenamefont{Bennett, Brassard,
  Crepeau, Jozsa, Peres, and Wootters}}]{Bennett-93}
\bibinfo{author}{\bibfnamefont{C.}~\bibnamefont{Bennett}},
  \bibinfo{author}{\bibfnamefont{G.}~\bibnamefont{Brassard}},
  \bibinfo{author}{\bibfnamefont{C.}~\bibnamefont{Crepeau}},
  \bibinfo{author}{\bibfnamefont{R.}~\bibnamefont{Jozsa}},
  \bibinfo{author}{\bibfnamefont{A.}~\bibnamefont{Peres}}, \bibnamefont{and}
  \bibinfo{author}{\bibfnamefont{W.}~\bibnamefont{Wootters}},
  \bibinfo{journal}{Phys.\ Rev.\ Lett.} \textbf{\bibinfo{volume}{70}},
  \bibinfo{pages}{1895} (\bibinfo{year}{1993}).

\bibitem[{\citenamefont{Nelson}(1977)}]{Nelson}
\bibinfo{author}{\bibfnamefont{R.~J.} \bibnamefont{Nelson}},
  \bibinfo{journal}{Appl.\ Phys.\ Lett.} \textbf{\bibinfo{volume}{31}},
  \bibinfo{pages}{351} (\bibinfo{year}{1977}).

\bibitem[{\citenamefont{Kukushkin et~al.}(1989)\citenamefont{Kukushkin, von
  Klitzing, Ploog, Kirpichev, and Shepel}}]{Klitzing-89}
\bibinfo{author}{\bibfnamefont{I.}~\bibnamefont{Kukushkin}},
  \bibinfo{author}{\bibfnamefont{K.}~\bibnamefont{von Klitzing}},
  \bibinfo{author}{\bibfnamefont{K.}~\bibnamefont{Ploog}},
  \bibinfo{author}{\bibfnamefont{V.}~\bibnamefont{Kirpichev}},
  \bibnamefont{and} \bibinfo{author}{\bibfnamefont{B.}~\bibnamefont{Shepel}},
  \bibinfo{journal}{Phys.\ Rev.\ B} \textbf{\bibinfo{volume}{40}},
  \bibinfo{pages}{4179} (\bibinfo{year}{1989}).

\bibitem[{\citenamefont{Chen et~al.}(1992)\citenamefont{Chen, Yang, Wilson, and
  Yang}}]{Wilson}
\bibinfo{author}{\bibfnamefont{J.}~\bibnamefont{Chen}},
  \bibinfo{author}{\bibfnamefont{C.}~\bibnamefont{Yang}},
  \bibinfo{author}{\bibfnamefont{R.}~\bibnamefont{Wilson}}, \bibnamefont{and}
  \bibinfo{author}{\bibfnamefont{M.}~\bibnamefont{Yang}},
  \bibinfo{journal}{Appl.\ Phys.\ Lett.} \textbf{\bibinfo{volume}{60}},
  \bibinfo{pages}{2113} (\bibinfo{year}{1992}).

\bibitem[{\citenamefont{Field et~al.}(1993)\citenamefont{Field, Smith, Pepper,
  Ritchie, Frost, Jones, and Hasko}}]{Field}
\bibinfo{author}{\bibfnamefont{M.}~\bibnamefont{Field}},
  \bibinfo{author}{\bibfnamefont{C.}~\bibnamefont{Smith}},
  \bibinfo{author}{\bibfnamefont{M.}~\bibnamefont{Pepper}},
  \bibinfo{author}{\bibfnamefont{D.}~\bibnamefont{Ritchie}},
  \bibinfo{author}{\bibfnamefont{J.}~\bibnamefont{Frost}},
  \bibinfo{author}{\bibfnamefont{G.}~\bibnamefont{Jones}}, \bibnamefont{and}
  \bibinfo{author}{\bibfnamefont{D.}~\bibnamefont{Hasko}},
  \bibinfo{journal}{Phys.\ Rev.\ Lett.} \textbf{\bibinfo{volume}{70}},
  \bibinfo{pages}{1311} (\bibinfo{year}{1993}).

\bibitem[{\citenamefont{Ciorga et~al.}(2000)\citenamefont{Ciorga, Sachrajda,
  Hawrylak, Gould, Zawadzki, Jullian, Feng, and Wasilewski}}]{Ciorga}
\bibinfo{author}{\bibfnamefont{M.}~\bibnamefont{Ciorga}},
  \bibinfo{author}{\bibfnamefont{A.}~\bibnamefont{Sachrajda}},
  \bibinfo{author}{\bibfnamefont{P.}~\bibnamefont{Hawrylak}},
  \bibinfo{author}{\bibfnamefont{C.}~\bibnamefont{Gould}},
  \bibinfo{author}{\bibfnamefont{P.}~\bibnamefont{Zawadzki}},
  \bibinfo{author}{\bibfnamefont{S.}~\bibnamefont{Jullian}},
  \bibinfo{author}{\bibfnamefont{Y.}~\bibnamefont{Feng}}, \bibnamefont{and}
  \bibinfo{author}{\bibfnamefont{Z.}~\bibnamefont{Wasilewski}},
  \bibinfo{journal}{Phys.\ Rev.\ B} \textbf{\bibinfo{volume}{61}},
  \bibinfo{pages}{R16315} (\bibinfo{year}{2000}).

\bibitem[{\citenamefont{Sprinzak et~al.}(2002)\citenamefont{Sprinzak, Ji,
  Heiblum, Mahalu, and Shtrikman}}]{Shtrikman1}
\bibinfo{author}{\bibfnamefont{D.}~\bibnamefont{Sprinzak}},
  \bibinfo{author}{\bibfnamefont{Y.}~\bibnamefont{Ji}},
  \bibinfo{author}{\bibfnamefont{M.}~\bibnamefont{Heiblum}},
  \bibinfo{author}{\bibfnamefont{D.}~\bibnamefont{Mahalu}}, \bibnamefont{and}
  \bibinfo{author}{\bibfnamefont{H.}~\bibnamefont{Shtrikman}},
  \bibinfo{journal}{Phys.\ Rev.\ Lett.} \textbf{\bibinfo{volume}{88}},
  \bibinfo{pages}{176805} (\bibinfo{year}{2002}).

\bibitem[{\citenamefont{Elzerman et~al.}(2003)\citenamefont{Elzerman, Hanson,
  Greidanus, van Beveren, Franceschi, Vandersypen, Tarucha, and
  Kouwenhoven}}]{Kouwen2003-1}
\bibinfo{author}{\bibfnamefont{J.}~\bibnamefont{Elzerman}},
  \bibinfo{author}{\bibfnamefont{R.}~\bibnamefont{Hanson}},
  \bibinfo{author}{\bibfnamefont{J.}~\bibnamefont{Greidanus}},
  \bibinfo{author}{\bibfnamefont{L.~W.} \bibnamefont{van Beveren}},
  \bibinfo{author}{\bibfnamefont{S.~D.} \bibnamefont{Franceschi}},
  \bibinfo{author}{\bibfnamefont{L.}~\bibnamefont{Vandersypen}},
  \bibinfo{author}{\bibfnamefont{S.}~\bibnamefont{Tarucha}}, \bibnamefont{and}
  \bibinfo{author}{\bibfnamefont{L.}~\bibnamefont{Kouwenhoven}},
  \bibinfo{journal}{Phys.\ Rev.\ B} \textbf{\bibinfo{volume}{67}},
  \bibinfo{pages}{R161308} (\bibinfo{year}{2003}).

\bibitem[{\citenamefont{Elzerman et~al.}(2004)\citenamefont{Elzerman, Hanson,
  van Beveren, Witkamp, Vandersypen, and Kouwenhoven}}]{Kouwen2004-1}
\bibinfo{author}{\bibfnamefont{J.}~\bibnamefont{Elzerman}},
  \bibinfo{author}{\bibfnamefont{R.}~\bibnamefont{Hanson}},
  \bibinfo{author}{\bibfnamefont{L.~W.} \bibnamefont{van Beveren}},
  \bibinfo{author}{\bibfnamefont{B.}~\bibnamefont{Witkamp}},
  \bibinfo{author}{\bibfnamefont{L.}~\bibnamefont{Vandersypen}},
  \bibnamefont{and}
  \bibinfo{author}{\bibfnamefont{L.}~\bibnamefont{Kouwenhoven}},
  \bibinfo{journal}{Nature} \textbf{\bibinfo{volume}{430}},
  \bibinfo{pages}{431} (\bibinfo{year}{2004}).

\bibitem[{\citenamefont{Xiao et~al.}(2004)\citenamefont{Xiao, Martin,
  Yablonovitch, and Jiang}}]{MingX}
\bibinfo{author}{\bibfnamefont{M.}~\bibnamefont{Xiao}},
  \bibinfo{author}{\bibfnamefont{I.}~\bibnamefont{Martin}},
  \bibinfo{author}{\bibfnamefont{E.}~\bibnamefont{Yablonovitch}},
  \bibnamefont{and} \bibinfo{author}{\bibfnamefont{H.~W.} \bibnamefont{Jiang}},
  \bibinfo{journal}{Nature} \textbf{\bibinfo{volume}{430}},
  \bibinfo{pages}{435} (\bibinfo{year}{2004}).

\end{thebibliography}
\end{document}